\documentclass{emulateapj} \usepackage{times}

\usepackage{color} 
\usepackage{epsfig}

\newcommand{\etal}{{et~al.}}

\newcommand{\be}{\begin{equation}}
\newcommand{\ee}{\end{equation}}
\newcommand{\bea}{\begin{eqnarray}}
\newcommand{\eea}{\end{eqnarray}}

\shorttitle{NEATM vs Thermophysical}
\shortauthors{Wright}
\journalinfo{}
\submitted{}

\begin{document}

\title{Comparing the NEATM with a Rotating, Cratered Thermophysical
Asteroid Model}

\author{Edward L. Wright}
\affil{UCLA Dept.\ of Physics \& Astronomy\\
PO Box 951547, Los Angeles CA 90095-1547}
\email{wright@astro.ucla.edu}

\clearpage

\vskip 2.0cm

\begin{abstract}

A cratered asteroid acts somewhat like a retroflector, sending light
and infrared radiation back toward the Sun, while thermal inertia in a
rotating asteroid causes the infrared radiation to peak over the
``afternoon'' part.  In this paper a rotating, cratered asteroid model
is described, and used to generate infrared fluxes which are then
interpreted using the Near Earth Asteroid Thermal Model (NEATM).  Even
though the rotating, cratered model depends on three parameters not
available to the NEATM (the dimensionless thermal inertia parameter and
pole orientation), the NEATM gives diameter estimates that are accurate
to 10\% RMS for phase angles less than $60^\circ$.  For larger phase
angles, such as back-lit asteroids, the infrared flux depends more strongly
on these unknown parameters, so the diameter errors are larger.
These results are still true for the non-spherical shapes typical of small
Near Earth objects.

\end{abstract}

\keywords{asteroids, size, infrared}

\section{Introduction}
\label{sec:intro}

Rotation causes a diurnal oscillation in the illuminating flux on a surface element
of an asteroid.  During the day, heat is conducted into the surface, while during
the night this heat is radiated.  The combination of this phase delayed conducted
heat and the direct heat from the Sun leads to a temperature maximum during the
afternoon on an asteroid, just as it does on the Earth.
Rotating models of planets \citep{wri76} and asteroids \citep{pet76}
which incorporate the effects of thermal inertia have been in use for
decades.  But these models do not show the peaking near zero phase angle
seen in real asteroids. This lack is addressed in the standard thermal 
model (STM) for asteroids \citep{leb86} by evaluating the flux at zero
phase angle and then applying a linear 0.01 mag/degree phase
correction.  This beaming of the infrared radiation toward the Sun reduces
the total reradiation, so in order to conserve energy the subsolar
temperature is computed by replacing the emissivity of the surface $\epsilon$
by $\epsilon\eta$, where the beaming correction
$\eta = 0.756$.  This approximately conserves energy,
but the STM is really an empirical fitting function rather than a physical model.
Infrared observations of Near Earth Objects (NEOs) analyzed using the STM
yielded inaccurate diameters for small, rapidly rotating NEOs seen at large
phase angles, so \citet{har98} developed the Near Earth Asteroid Thermal
Model (NEATM).  In the NEATM the beaming correction $\eta$ is an adjustable
parameter, and the infrared flux is evaluated by integrating the emission over
the asteroid surface seen from the actual position of the observer.  The observed
color temperature is matched by adjusting $\eta$, and then the NEATM specifies
the average surface brightness of the object, so the observed flux implies a
diameter.

Neither the STM nor the NEATM has a physical explanation for the
beaming effect, but \citet{han77} provides one using a cratered
asteroid model.  If an asteroid is covered with craters, the peak
temperature on the surface at the sub-solar point will be higher.
Furthermore, the flux at zero phase angle from areas near the limb will
be much higher than in either STM or the NEATM, since while the surface
is a mixture of illuminated and shadowed areas, the illuminated areas
are closer to facing the Sun, and an observer at zero phase angle sees
only the lighted areas.  The lack of visible shadows at zero phase
angle cause a peak in the emission at opposition.  In order to get
enough shadowing, it is necessary to have a substantial amount of
concavity in the surface, which could be due to a porous granular
surface, or to the craters considered by \citet{han77}.

Cratering and thermal inertia have been combined into a single
thermophysical model by \cite{lag96}.  The thermophysical model used in this
paper has been developed independently, but is very similar to the
\cite{lag96} model with the crater covered fraction set to 100\%,  and
with depth to width ratio given by
$(1-\cos\theta_{max})/(2\sin\theta_{max}) = 0.2$ for $\theta_{max} =
45^\circ$.

\section{Rotating Cratered Model}
\label{sec:rc}

\begin{figure}[tbh]
\plotone{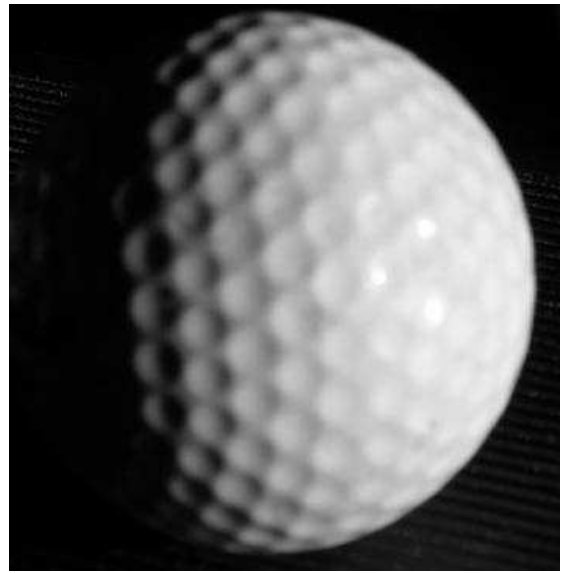}
\caption{Physical analog of the cratered asteroid model, seen at $\sim 60^\circ$
phase angle.  In the actual computational model the flat areas between the craters are
eliminated.
\label{fig:golf}}
\end{figure}

\begin{figure}[t]
\plotone{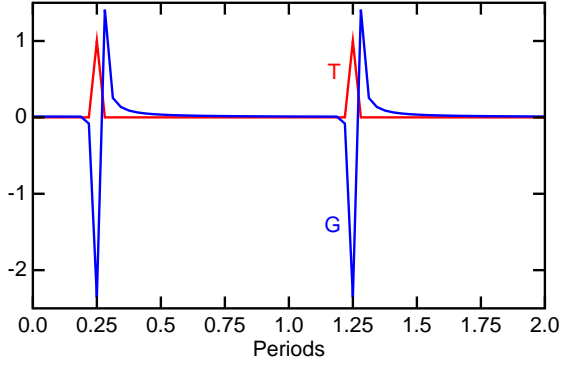}
\caption{Heat flow into the surface ($G$) for a triangle temperature forcing function $T$.
Computed with 32 samples per period.  Two full cycles are shown.
\label{fig:G}}
\end{figure}

\begin{figure}[tbh]
\plotone{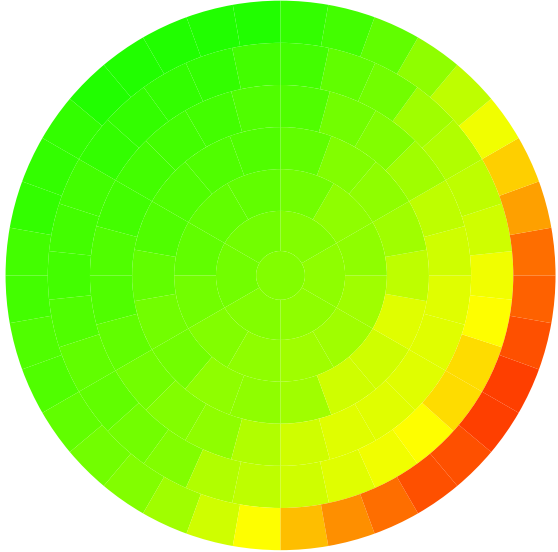}
\caption{Temperature distribution in the late afternoon for a crater at $4^/circ$ 
latitude when the sub-solar latitude is $30^\circ$.  The thermal inertia 
parameter is $\Theta = 1$.  The Sun is setting in the Northwest, so the 
Southeast rim of the crater is still sunlit.
\label{fig:frame}}
\end{figure}

\begin{figure}[tbh]
\plotone{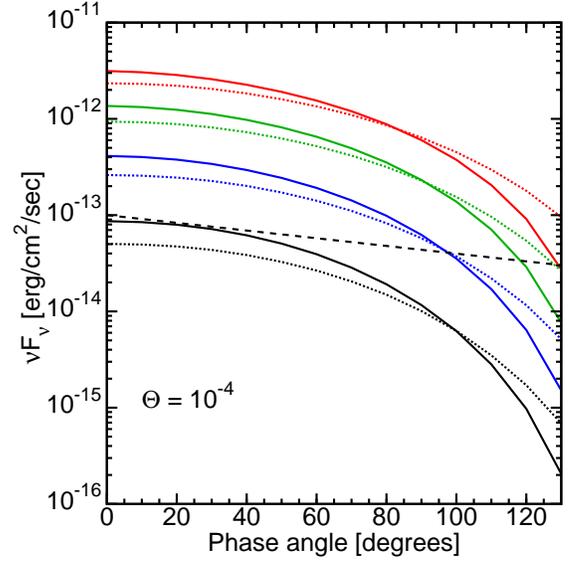}
\caption{Infrared phase curves for an asteroid 1.4 AU from the Sun
in the passbands of the IRAC on the Spitzer Space Telescope, 
3.6, 4.5, 5.6 \& 8 $\mu$m from bottom to top.  The solid curves
show a cratered model with maximum slope $\theta_{max} = 45^\circ$,
while the dotted curves show an uncratered model.  The dashed
line shows the 0.01 mag/degree phase curve of the STM.
\label{fig:phase}}
\end{figure}

\begin{figure*}[tb]
\plotone{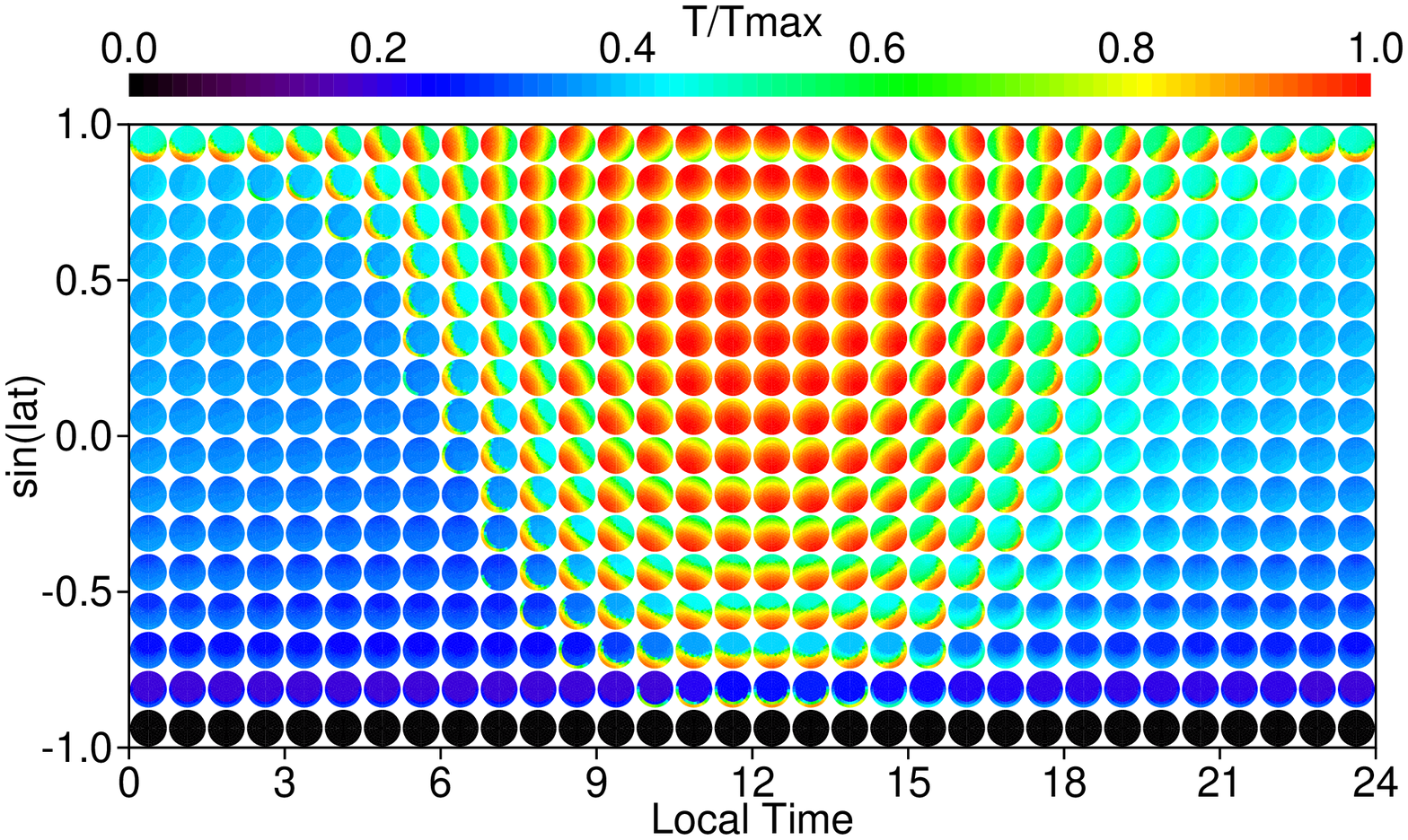}
\caption{Temperature distribution as a function of latitude and local
time for a rotating cratered model with thermal inertia parameter
$\Theta = 0.1$ and maximum crater slope $\theta_{max} = 45^\circ$.
The sub-solar latitude is $30^\circ$.
\label{fig:X0p1}}
\end{figure*}

\begin{figure*}[tb]
\plotone{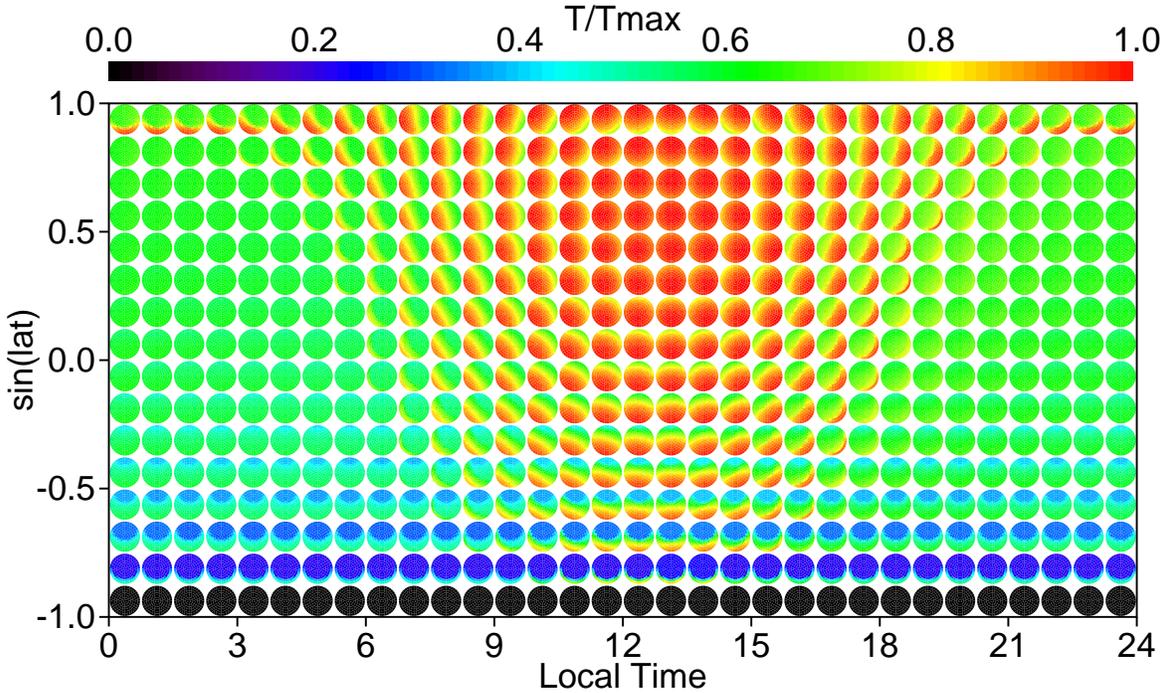}
\caption{Temperature distribution as a function of latitude and local
time for a rotating cratered model with thermal inertia parameter
$\Theta = 1$ and maximum crater slope $\theta_{max} = 45^\circ$.
The sub-solar latitude is $30^\circ$.   The circle just above the equator
at local time 17:37 is blown up in Figure \ref{fig:frame}.
\label{fig:X1}}
\end{figure*}

\begin{figure*}[tb]
\plotone{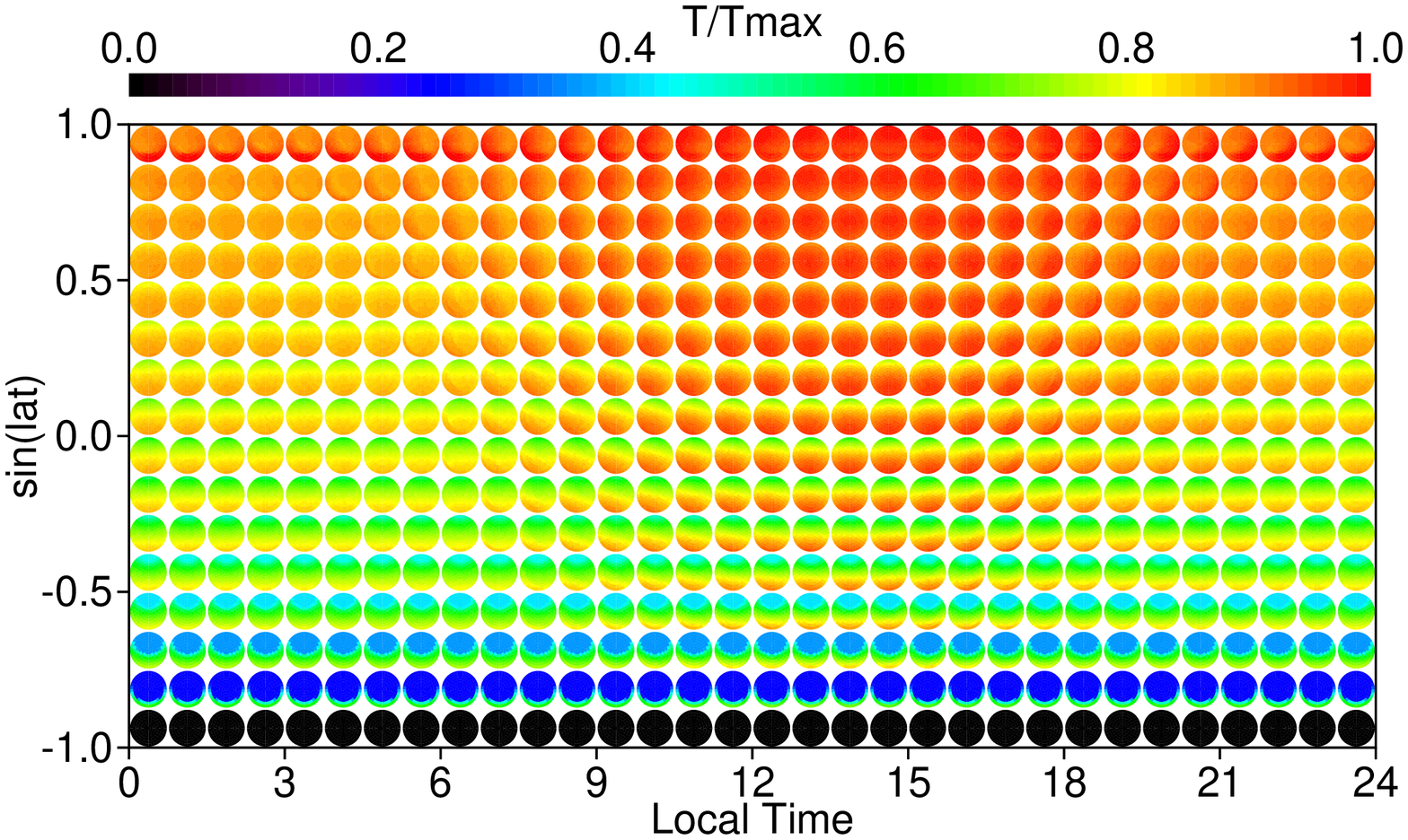}
\caption{Temperature distribution as a function of latitude and local
time for a rotating cratered model with thermal inertia parameter
$\Theta = 10$ and maximum crater slope $\theta_{max} = 45^\circ$.
The sub-solar latitude is $30^\circ$.
\label{fig:X10}}
\end{figure*}

\begin{figure}[t]
\plotone{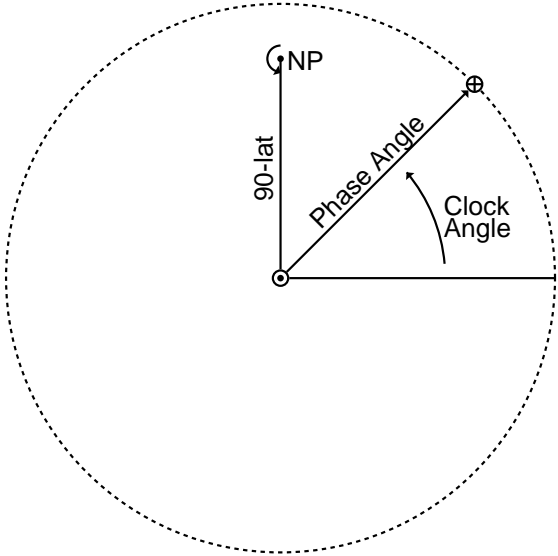}
\caption{Definition of the clock angle.  $\odot$ marks
the sub-solar point, and $\oplus$ marks the sub-observer
point.  The angle between the pole and the sub-solar
point is $90-\beta_\odot$.
\label{fig:angles}}
\end{figure}

\begin{figure}[t]
\plotone{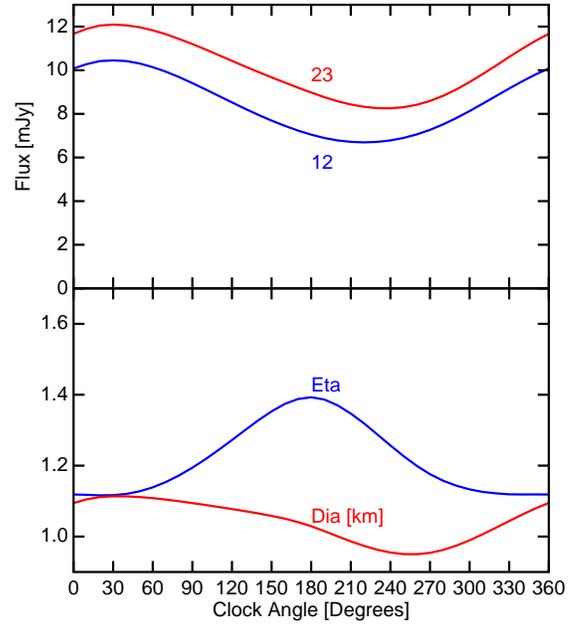}
\caption{Flux in the WISE 12 and 23 $\mu$m bands for a 1 km 
asteroid 1 AU from the Earth and 1.4 AU from the Sun, observed
at $50^\circ$ phase angle.  The sub-solar latitude is $30^\circ$
and the thermal inertia parameter is $\Theta = 1$.  The clock
angle rotates the line of sight around the Sun line while
keeping a constant phase angle.  For $0^\circ$ clock angle
the observer is East of the Sun, giving an afternoon view.
For $90^\circ$ clock angle the observer is over $80^\circ$
latitude at noon.
These fluxes lead to a beaming parameter $\eta$ and 
calculated diameter $D$ using the NEATM.
\label{fig:vs50}}
\end{figure}

\begin{figure}[b]
\plotone{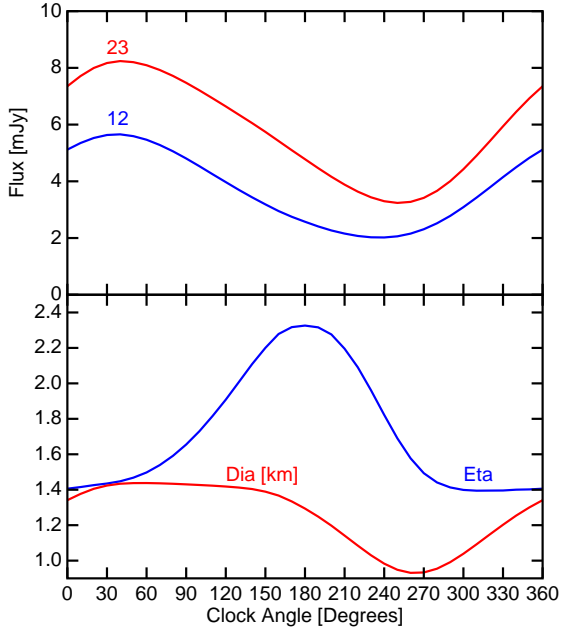}
\caption{As in Figure \ref{fig:vs50}, but
at $90^\circ$ phase angle.   For $0^\circ$ clock angle
the observer is over the equator at 6 PM.
For $90^\circ$ clock angle the observer is over $60^\circ$
latitude at midnight.
\label{fig:vs90}}
\end{figure}

\begin{figure*}[t]
\plotone{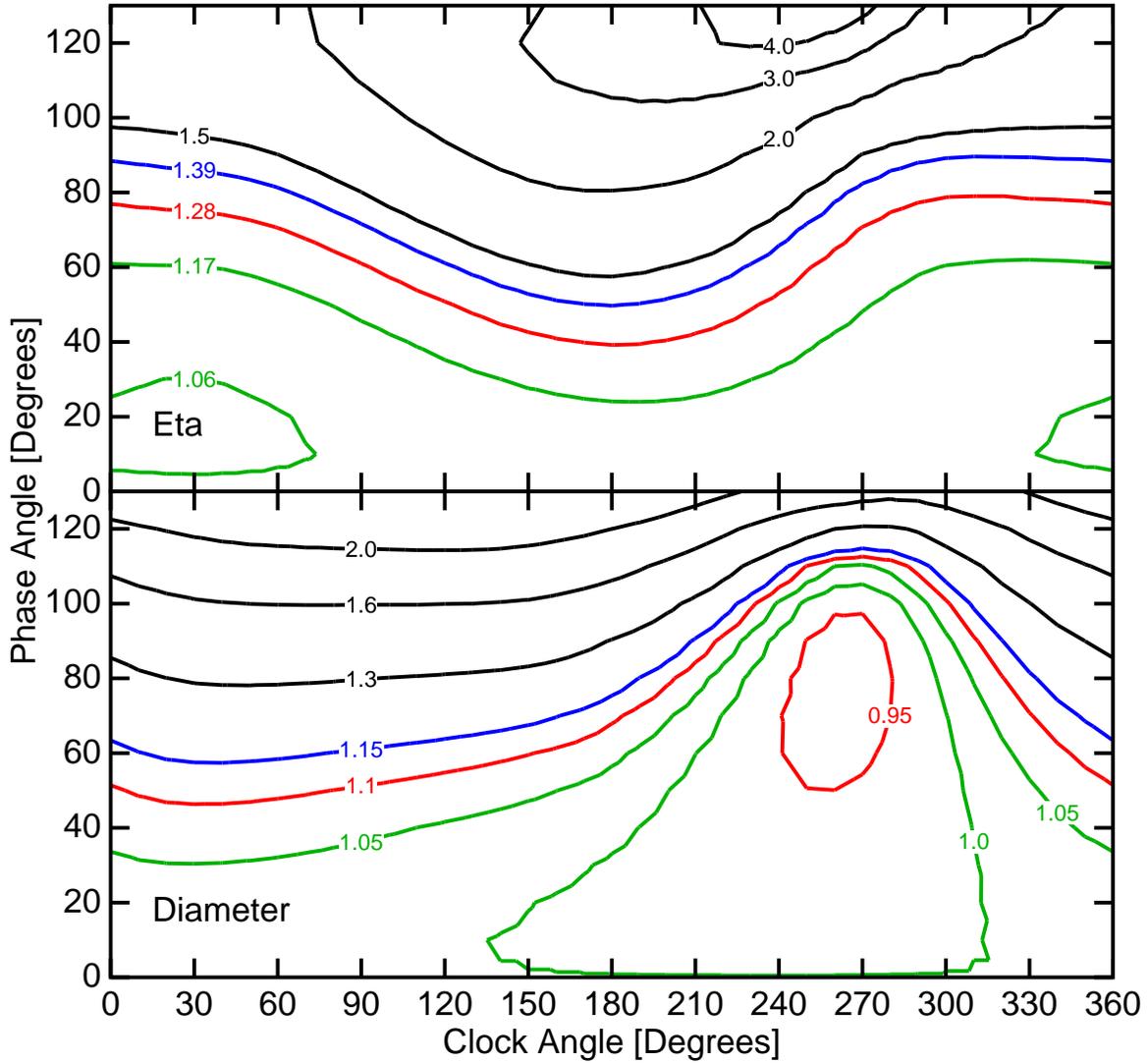}
\caption{The beaming parameter $\eta$ and the fitted diameter as
a function of the phase and clock angles for the asteroid in
Figure \ref{fig:vs50}.
\label{fig:eta-dia}}
\end{figure*}

\begin{figure}[t]
\plotone{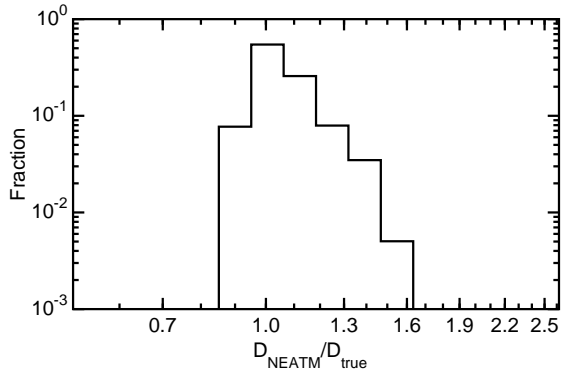}
\caption{Histogram showing the frequency of various diameter discrepancies
when using the NEATM to analyze fluxes from the rotating cratered model, 
assuming spherical asteroids.
\label{fig:hist}}
\end{figure}

\begin{figure*}[t]
\plotone{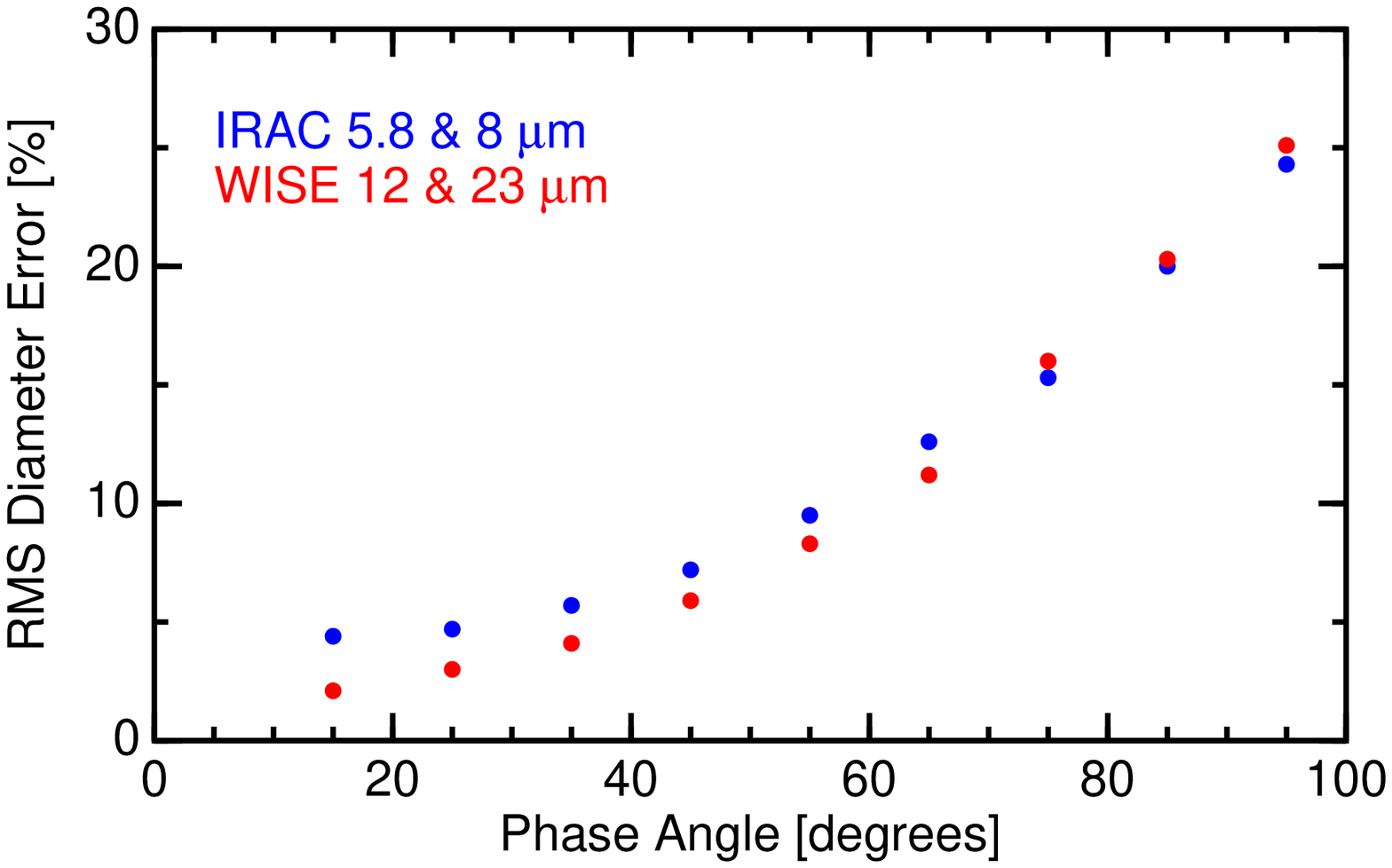}
\caption{RMS diameter errors for a sample of 576 NEOs that could be
observed by the Spitzer Space Telescope in Cycle 4.  While the
distance from the the Sun and phase angle are held fixed at the
actual values, fluxes were simulated for Monte Carlo asteroids
with random thermal inertias uniform in pole position and
$\log\Theta$ in the range $0.1 < \Theta < 10$.  Fluxes at
the WISE
12 \& 23 $\mu$m bands gave NEATM RMS diameter
errors shown by the lower dots.  The IRAC 5.8 and 8 $\mu$m
fluxes give the upper dots.
\label{fig:sigma}}
\end{figure*}

\begin{figure}[t]
\plotone{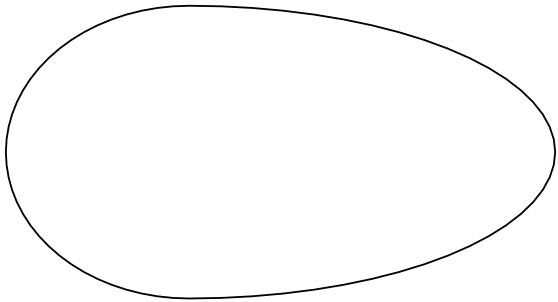}
\caption{The egg shaped example.  The figure is rotationally symmetric
about the horizontal axis, while the object rotates about a vertical axis.
\label{fig:egg-shape}}
\end{figure}

\begin{figure}[t]
\plotone{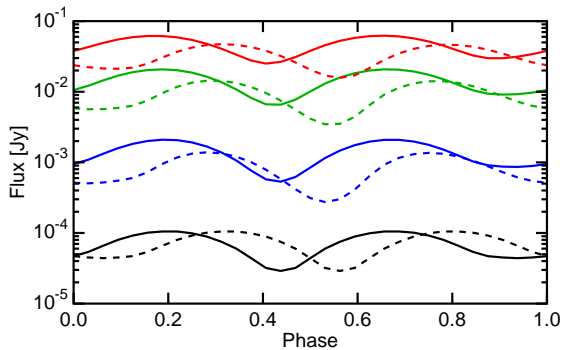}
\caption{Lightcurves for the egg shaped asteroid seen at $45^\circ$ phase
angle with the rotation axis normal to the Earth-Sun-asteroid plane.
The solid curves show viewing the afternoon side, while the dashed
curves show the morning side.
Black is optical, blue is 3.6 $\mu$m, green is 5.8 $\mu$m, and red is
12 $\mu$m.
\label{fig:lightcurve}}
\end{figure}

\begin{figure}[t]
\plotone{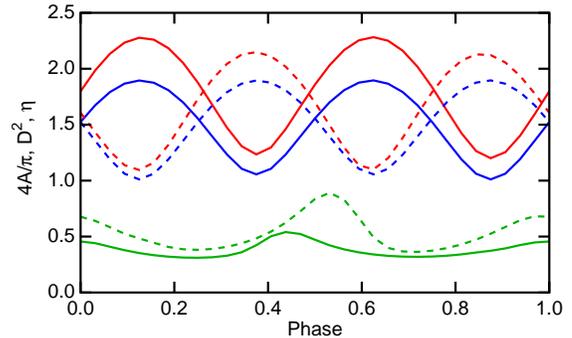}
\caption{The variations of the projected area in blue, the
square of the NEATM derived diameter in red, and the
NEATM $\eta$ parameter in green.
\label{fig:A-D-eta}}
\end{figure}

\begin{figure}[t]
\plotone{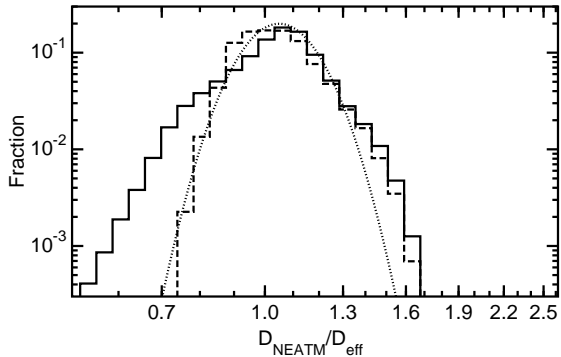}
\caption{Histogram showing the frequency of various diameter discrepancies
when using the NEATM to analyze fluxes from the rotating cratered 
egg-shaped model.  The solid histogram shows the distribution of diameter
errors derived from single observations at a random rotational phase.
The dashed histogram shows the distribution of diameter errors
derived from the average over a rotation.  The dotted curve shows
a Gaussian with a 10\% standard deviation.
\label{fig:hist3}}
\end{figure}

\begin{figure*}[t]
\plotone{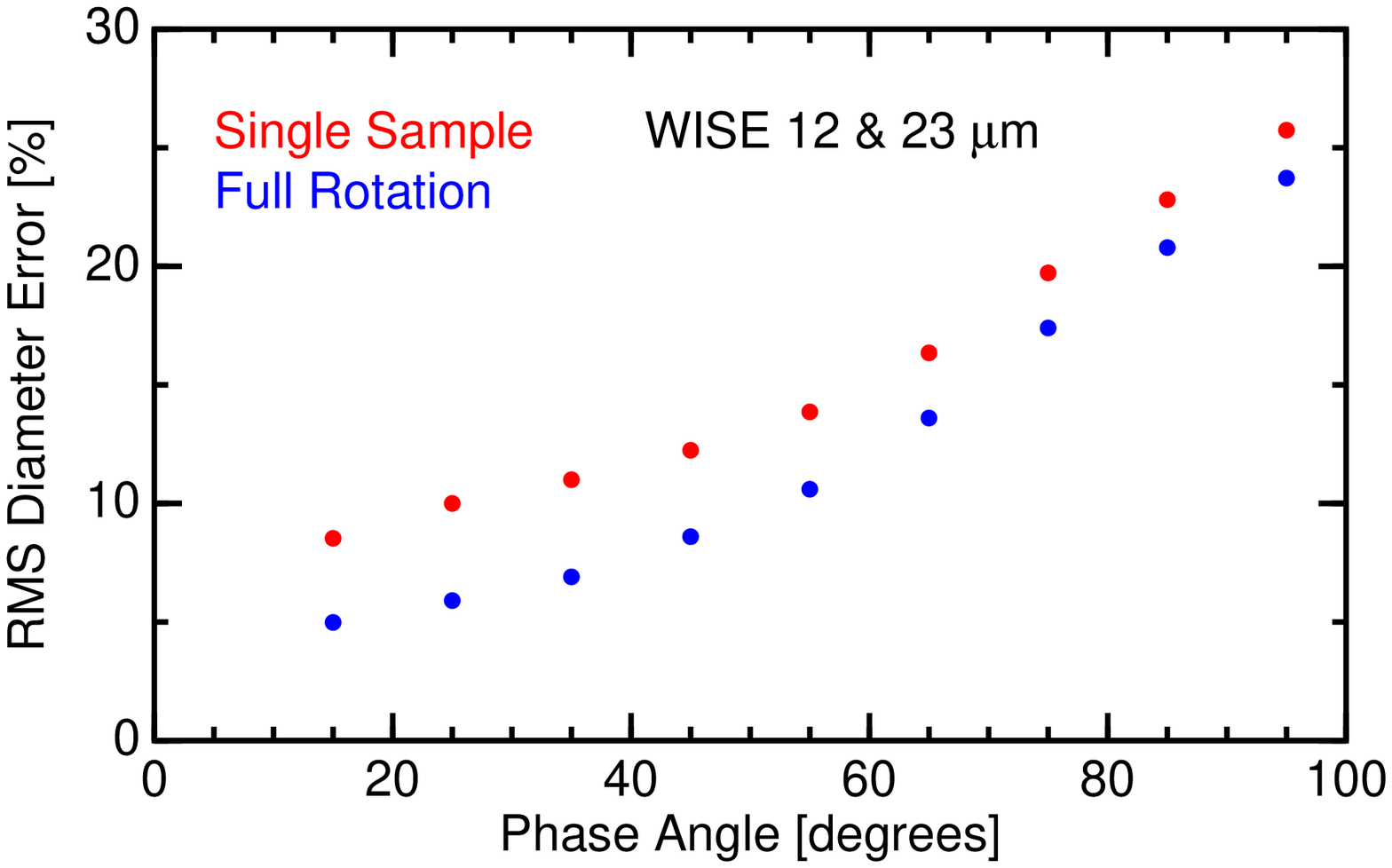}
\caption{RMS diameter errors for a sample of 576 NEOs that could be
observed by the Spitzer Space Telescope in Cycle 4.  While the
distance from the the Sun and phase angle are held fixed at the
actual values, fluxes were simulated for Monte Carlo asteroids
with random thermal inertias uniform in pole position and
$\log\Theta$ in the range $0.1 < \Theta 10$.  Fluxes at
the WISE
12 \& 23 $\mu$m bands were used in the  NEATM. 
RMS diameter errors when a single observation is made of each asteroid
give the upper dots, while errors after averaging the NEATM diameter 
give the lower dots.
\label{fig:sigmarot}}
\end{figure*}

In the rotating cratered asteroid model, heat is conducted 
vertically into and out of the surface, but not horizontally.
The facets in a given crater can see each other and see the Sun,
but there is no radiative transfer of heat from one crater to 
another.  Thus the problem of finding the temperature
distribution on the asteroid breaks up into many small
problems of finding the temperature {\it vs.} time of each facet.

The equation describing heat conduction in the surface layer is
\be
C \rho {{\partial T} \over {\partial t}} =
\kappa {{\partial^2 T} \over {\partial z^2}}
\ee
where $\kappa$ is the thermal
conductivity, $\rho$ is the density, and $C$ is the specific heat per
unit mass.  The heat flow into the surface is
$\kappa {{\partial T} \over {\partial z}}$.
Letting $\zeta = z/\kappa$, then the heat flow is
${{\partial T} \over {\partial \zeta}}$ and the
heat conduction equation is
\be
\kappa\rho C {{\partial T} \over {\partial t}} =
{{\partial^2 T} \over {\partial \zeta^2}}
\ee
which depends only on
the thermal inertia $\Gamma = \sqrt{\kappa\rho C}$.
The units of $\Gamma$ are 
$[E/(t(T/L)L^2) M/L^3 E/(TM)]^{1/2} = E/(T L^2 \sqrt{t})$.
In \citet{wri76} the value of $\Gamma = 0.006\;
\mbox{cal/cm$^2$/K/sec$^{1/2}$}$ was used for Mars.
In more modern units this is $251\;\mbox{J/m$^2$/K/sec$^{1/2}$}$.
\citet{har06} gives values for $\Gamma$ of 10-20 for
main belt asteroids, 50 for the Moon, 150 \& 350 for the NEOs Eros
\& Itokawa, and $2500\;\mbox{J/m$^2$/K/sec$^{1/2}$}$ for bare rock.

Putting in the solar heating and thermal radiation, I get
\be
{{(1-A)L_\odot}\over{4\pi R^2}} \max(0,\cos\theta) - \epsilon\sigma T^4
= {{\partial T} \over {\partial \zeta}}
\ee
with $\theta = \Omega t$ where $\Omega$ is the rotational
frequency of the asteroid, $R$ is the distance of the asteroid from the Sun,
$L_\odot$ is the solar luminosity, $A$ is the albedo, $\epsilon$ is
the emissivity in the thermal infrared, and $\sigma$ is the
Stephan-Boltzmann constant.  This assumes the special case of
an equatorial region with the sub-solar latitude ($\beta_\odot$) of zero.  In
general the $\max(0,\cos\theta)$ would be replaced by either the
cosine of the angle between the Sun and the surface normal, or
zero if the facet is shadowed on the Sun is below the horizon.

The unit of temperature in the code is
\be
T_\circ = \left({{(1-A)L_\odot}\over{4\pi\epsilon\sigma R^2}}\right)^{1/4}
\ee
which is the equilibrium temperature of a surface oriented toward the Sun.
Let $y = T/T_\circ$, giving
\be
\epsilon\sigma T_\circ^3 [\max(0,\cos\theta) - y^4] =
{{\partial y} \over {\partial \zeta}}
\ee
and
\be
\kappa\rho C \Omega {{\partial y} \over {\partial \theta}} =
{{\partial^2 y} \over {\partial \zeta^2}}
\ee
Redefine the depth variable again using 
$w = \epsilon\sigma T_\circ^3 \zeta = z/H$
where $H$ is the distance such that a temperature
gradient of $T_\circ/H$ sets up a conductive flux equal
to the radiative flux $ \epsilon\sigma T_\circ^4$.
These equations are now
\be
\max(0,\cos\theta) - y^4 = {{\partial y} \over {\partial w}}
\ee
and
\be
{{\kappa\rho C \Omega}\over{(\epsilon\sigma T_\circ^3)^2}}
{{\partial y} \over {\partial \theta}} =
{{\partial^2 y} \over {\partial w^2}}.
\ee

The coefficient
\be
\Theta = {{\sqrt{\kappa\rho C \Omega}}\over{\epsilon\sigma T_\circ^3}} =
{{\Gamma \sqrt{\Omega R^3}} \over
     {(\epsilon \sigma)^{1/4}[(1-A)L_\odot/4\pi]^{3/4}}}
\ee
is a dimensionless measure of the importance of thermal inertia
on the temperatures.  It is about $\Theta = 1.5$ for Mars using
the parameters of \citet{wri76}.

Now let $y = \sum_n y_n \exp(in\theta-k_n w)$.  The solutions to the
equation
\be
\Theta^2 {{\partial y} \over {\partial \theta}} =
{{\partial^2 y} \over {\partial w^2}}
\ee
must have $in \Theta ^2 = k_n^2$ so $k_n = \Theta \sqrt{n/2}(1+i)$ for $n > 0$.
For negative $n$, $k_n = \Theta \sqrt{|n|/2}(1-i)$ must be taken
to guarantee the solution is damped toward $z = \infty$.

The physical length scale given by $H/\Theta = \Gamma/(C\rho\sqrt{\Omega})$.
Since $C\rho \approx 10^6\;\mbox{J/m$^3$/K}$ and $\Omega \approx 10^{-4}
\;\mbox{rad/sec}$, the length scale is smaller than 10 cm.  Thus the
assumption of no horizontal heat conduction is reasonable for craters
$> 1\;\mbox{m}$.

The heat flow into the surface is
\be
{{\partial y} \over {\partial w}} =
-\sum_n y_n k_n \exp(in\theta).
\ee
Now take $y(\theta,0)$ as a triangle wave centered at $\theta = 0$ with
$y = 1$, and linearly sloping down to $y = 0$ at
$\theta = \pm 2\pi/N$.  This gives
\bea
y_n & = & \pi^{-1} \int_0^{2\pi/N} \cos(n\theta) (1-N\theta/2\pi) d\theta
\nonumber \\
& = & {{N(1-\cos(2\pi n/N))}\over{2\pi^2 n^2}}
\eea
The heat flow into the surface is $\Theta G(\theta)$ with
\be
G(\theta) = \sum_{n=1}^\infty {{N(1-\cos(2\pi n/N))}\over{2\pi^2 n^2}}
\sqrt{2n} (\sin n\theta - \cos n\theta)
\ee
The value of this $G(\theta)$ at $\theta = 2\pi m/N$ defines
the vector $G$ used in code.
The average of this over bins of width $2\pi/N$ in $\theta$
centered $\theta = 2\pi m/N$ is:
\bea
\overline{G(m)} & = & \sum_{n=1}^\infty {{N^2(1-\cos[2\pi n/N])\sin[\pi n/N]}
\over{\sqrt{2}\pi^3 n^{2.5}}}
\nonumber \\
& \times & \left(\sin\left[{{2\pi nm}\over{N}}\right] -
\cos\left[{{2\pi nm}\over{N}}\right]\right)
\eea
Figure \ref{fig:G} shows the function $\overline{G(m)}$ for $N = 32$.

The mutual irradiation of the crater facets introduces a coupling
matrix.  The contribution of facet $j$ to facet $i$ goes like
$\pi^{-1} \Delta \Omega_j R^2 \cos(\theta_j) \cos(\theta)/d^2_{ij}$
where $\theta_i = \theta_j$ are the angle of emission and incidence,
$\Delta\Omega_j$ is the solid angle of the facet on the spherical
cap crater,
and $d_{ij} = 2R\cos(\theta_j)$ is the distance between the facets.
Thus the coupling is just $\Delta \Omega/4\pi$.
The total light falling on a facet is given by
\be
S_i = D_i + A \sum_j {{\Delta \Omega_j}\over{4\pi}} S_j = D_i +
{{A \Omega_{crater}}\over{4\pi}} \langle S \rangle.
\ee
where $D_i$ is the direct solar flux, and $S_i$ is the total flux on
a facet.
Averaging this equation gives
\be
\langle S \rangle = \langle D \rangle +
{{A \Omega_{crater}}\over{4\pi}} \langle S \rangle
\ee
so
\be
\langle S \rangle = {{\langle D \rangle}\over{1 - A )f_c}}
\ee
where $f_c$ is $\Omega_{crater}/4\pi$, the fraction of the
sphere included in the spherical cap craters.  For the
$\theta_{max} = 45^\circ$ used here, $f_c = (1-\cos\theta_{max})/2 = 0.15$.
Therefore
\be
S_i = D_i + {{\langle D \rangle A f_c}\over
{1 - A f_c}}
\ee

For the calculations reported here, the craters were divided into
127 facets, consisting of a central circle surrounded by rings
of 6, 12, \ldots, 36 square facets.  Since the resulting facet
size of $2\theta_{max}/13 \approx 7^\circ$ was fairly coarse,
the direct insolation $D_i$ was computed as $\mu_i f_v$ times
$F_\odot$, where $f_v$ is the fraction of the facet that is
visible from the Sun, and $\mu$ is the cosine of the angle
between the surface normal and the Sun.  
The visible fraction is computed using
a finer pixelization of the sphere, HEALpix  \citep{gor05} with 49,152 pixels.
Figure \ref{fig:frame} shows the facet structure with a crater.

In the thermal infrared, the mutual visibility of the facets couples
their temperatures together.  The total infrared flux falling on a
facet is
\be
H_i = \sum_j {{\epsilon\sigma T_j^4 \Delta \Omega_j}
\over {4\pi}} + (1-\epsilon) f_c \langle H \rangle
\ee
Averaging this equation gives
\be
H_i = <H> = {{\epsilon\sigma \langle T^4 \rangle f_c}
\over {1 - (1-\epsilon) f_c}}
\ee

Remembering that the unit of flux is both $(1-A)F_\odot$ and
$\epsilon\sigma T_\circ^4$, and that only a fraction $\epsilon$
of the incident heat $H$ is absorbed, we get a power balance equation
\bea
y_{ij}^4 & = & (f_v\mu)_{ij} + \frac{A f_c}{1-A f_c} n^{-1} \sum_j  (f_v\mu)_{ij} 
\nonumber \\
& + & \frac{\epsilon f_c}{1-(1-\epsilon) f_c} n^{-1} \sum_j y_{ij}^4
\nonumber \\
& + & \Theta \sum_k  y_{ik} G_{[(k-j) \mbox{mod} N]} 
\eea
where $y_{ij}$ is the temperature of the $i^{th}$ facet at the
$j^{th}$ time, and there are $n$ facets and $N$ times.
This is a set of $n \times N = 127 \times 32 = 4064$ coupled
non-linear equations in 4064 variables.  Fortunately the
facet to facet coupling is weak and can be handled by iteration,
so the task of solving for the temperatures is not too onerous.

There is little point in using extremely fine subdivisions of the 
asteroid's surface.   The brightness temperatures of Mars
were calculated by \citet{wri76} using only 6 latitudes.
In the calculations reported here, 16 latitudes
were used, and the rotation period was divided in 32 time
steps.  This gives 65,024 temperatures to be found.  These
temperature pattern are plotted in Figures \ref{fig:X0p1},
\ref{fig:X1} \& \ref{fig:X10} for $\Theta$ values of 0.1, 1 \& 10
with sub-solar latitude of $30^\circ$.

\subsection{Observed Flux Calculation}

Given the temperature distribution, the observed flux is found
by integrating over the surface of the asteroid.  For a given
frequency $\nu$, the quantity $x_\circ = h\nu/kT_\circ$
is found.  Then the observed infrared flux is found using
\bea
F_\nu & = & \frac{2\epsilon h\nu^3}{c^2} 
\sum_{\beta,j} \max(\mu_f(\beta,j),0) \frac{d\Sigma(\beta,j)}{D^2}
\nonumber \\
& \times & \frac{\sum_i (f_v\mu)_{ij\beta}(\exp(x_\circ/y_{ij\beta})-1)^{-1}}
{\sum_i (f_v\mu)_{ij\beta}}
\eea
where $d\Sigma(\beta,j)$ is the surface area of the asteroid in the bin
at latitude $\beta$ and longitude given by $j$, D is the distance to
the observer, $\mu_f$ is the cosine of angle between the normal to the
surface and the line of sight, and $(f_v\mu)_{ij\beta}$ is the fraction of
the $i^{th}$ facet visible by the observer times the cosine of the
angle between the facet normal and the line of sight.
The observed bolometric optical flux is
\bea
F_{opt} & = & \sum_{\beta,j} \max(\mu_f(\beta,j),0) \frac{d\Sigma(\beta,j)}{D^2}
\nonumber \\
& \times & \frac{\sum_i (f_v\mu)_{ij\beta} A S_{ij\beta}/\pi}{\sum_i (f_v\mu)_{ij\beta}}
\eea
This can be multiplied by $F_\nu(\odot)/F_{bol}(\odot)$ to give the reflected
optical spectrum.

In this paper the latitudes are uniformly spaced in $\sin\beta$, so
$d\Sigma$ is a constant for a sphere.  The assumption that the asteroid is spherical
only enters into $d\Sigma$, so any convex shape for an asteroid can be
accommodated merely by changing the weights $d\Sigma$ going into the
flux sum.  It is important to remember that the angles $\beta$ and $\lambda$
refer to the surface normal vector, not the vector from the center to a surface
element.  Thus at the end of the major axis of a $2:1:1$ ellipsoid, $d\Sigma$
is a minimum, 16 times smaller than the value at the end of the minor
or intermediate axes.

\section{NEATM Fitting}
\label{sec:NEATM}

The rotating cratered asteroid model described model has been used to
predict infrared fluxes for NEOs, and then these fluxes have been used
in the NEATM to find the beaming parameter $\eta$ and the diameter
$d$.  For a given observation, the distance to the Sun and the
phase angle are known, but there are still two angles and the thermal
inertia parameter $\Theta$ that need to be specified.
Figure \ref{fig:angles} shows the definition of the two angles, which
are the sub-solar latitude and the clock angle.
Examples are shown in Figures \ref{fig:vs50} and \ref{fig:vs90}
for a 1 km diameter asteroid 1.4 AU from the Sun, with albedo $A = 0.1$
and emissivity $\epsilon = 0.9$.  At a $50^\circ$ phase angle, the
errors in the NEATM calculated diameters are small, but for $90^\circ$
the maximum error increases by a factor of 4.  These calculations have
been done using the planned 12 \& 23 $\mu$m bands of the Widefield
Infrared Survey Explorer (WISE,  \citep{amy06}) which is scheduled for
launch in 2009.

\section{NEATM Accuracy}
\label{sec:disc}

The NEATM can reproduce the model diameters for quite well for phase
angles $\alpha < 60^\circ$.   Figure \ref{fig:eta-dia} shows that the
errors are small for any clock angle as long as the phase angle is
smaller than $60^\circ$.  This conclusion does not depend strongly on
either the sub-solar latitude or the thermal inertia parameter
$\Theta$.  To show this for a representative sample of real
observations, Monte Carlo simulations of NEO observations from a
Spitzer Space Telescope proposal (SNEAS, PI Eisenhardt).  576 NEOs were
found to be observable by Spitzer during Cycle 4, with good
signal-to-noise ratio in the IRAC 5.8 \& 8 $\mu$m bands.  Only the
distance to the Sun and the phase angle were taken from this
observation table.  Then fluxes were computed for a random distribution
of pole positions and thermal inertias.  The pole positions were chosen
uniformly in $4\pi$ steradians, and the thermal inertias were chosen
uniformly in the logarithm in the range $0.1 < \Theta < 10$.   To choose
a random pole position one picks $\sin\beta_\odot$ uniform in
$[0-1]$ and the clock angle uniform in $[0-2\pi]$.  For each
of the 576 objects 30 different choices of pole position and $\Theta$
were analyzed.  The  fluxes were then analyzed using the NEATM
to derive $\eta$ and a diameter.  The RMS diameter errors,
binned by phase angles, are shown in Figure \ref{fig:sigma}.  Both the
WISE 12 \& 23 $\mu$m bands, and the Spitzer IRAC 5.8 \& 8 $\mu$m bands
give data that work well with the NEATM.

When the NEATM breaks down at large phase angles, the estimated
diameter is usually too large.  Figure \ref{fig:hist} shows the distribution
of the errors for the Monte Carlo observations in Figure \ref{fig:sigma}.
The distribution is clearly positively skewed.

Figure \ref{fig:sigma} shows that the NEATM works reasonably well for moderate
phase angles when compared to a more complete thermophysical model.
But to show that the NEATM works in the real world one needs
comparisons to real objects with size determined by radar or spacecraft
imaging.  \citet{har02} find that NEATM diameter errors average less than
10\% for phase angles less than $60^\circ$, but the number of objects
in the comparison was quite small.

\section{Asphericity}

All of the previous caculations have assumed spherical objects, but small NEOs
usually have quite aspherical shapes.  An example of a non-spherical shape is
the egg-shaped object seen in Figure \ref{fig:egg-shape}.  This is half of a 2.5:1:1 
ellipsoid joined to half of a 1.25:1:1 ellipsoid.  If this object is 1 AU from the Earth,
and 1.414 AU from the Sun, with phase angle of $45^\circ$
and a sub-solar latitude of $0^\circ$, thermal
inertia parameter $\Theta = 1$, albedo of 0.1
and emissivity of 0.9, one gets the lightcurves seen in Figure \ref{fig:lightcurve}
for clock angles of $0^\circ$ and $180^\circ$.  
The flux  normalization applies to a 1 km short axis and a 1.875 km long axis.
The optical lightcurve amplitude is 1.4 mag
peak to peak while the 23 $\mu$m amplitude is 0.84 to 0.93 mag.  These amplitudes are
larger than the 1.875:1 variation in the projected area for this shape.
The NEATM applied to the 12 and 23 $\mu$m data gives the $\eta$ and $D$ values
plotted in Figure \ref{fig:A-D-eta} along with 
the projected area as a function of rotational phase.
The diameter from the NEATM tracks the projected area fairly well, and there is a
definite variation of the beaming parameter $\eta$ with rotational phase.

The calculations of the accuracy of the NEATM for the Spitzer sample of 576 NEOs
have been repeated for the egg-shaped asteroid shown in Figure \ref{fig:egg-shape}. 
A non-spherical shape introduces another parameter, the rotational phase, that must
be either treated as a random variable or integrated over.  Figure \ref{fig:hist3}
compares NEATM diameters to the true ``diameter'' of the egg.  There are many ways
to define the true diameter:  the diameter of the sphere with the same volume of the
egg is 1.233 km, while the sphere with the same surface area as the egg has a diameter
of 1.275 km.  Since the NEATM is trying to estimate the projected area of an object it
seems reasonable to use the equivalent area diameter of 1.275 km as the reference.

Even a single sample at a random rotational phase gives a reasonable diameter
estimate:  the ratio of the NEATM to equivalent area sphere has a median and
interquartile range of $1.042^{+0.075}_{-0.096}$.  Since the interquartile range
only contains 50\% of the sample, these errors should be considered to be 
standard errors instead of standard deviations.  One can do better
using the NEATM diameter averaged over the rotation period.  This gives 
a median ratio and interquartile range of $1.019^{+0.082}_{-0.072}$.
The improvement coming from better sampling of the rotational phase
is fairly small, even though the egg shape considered here gives high peak to peak 
amplitudes up to  1.4 mag in the optical and short infrared wavelengths.
The small amplitude in diameters is due to several factors: the lightcurve
amplitude is lower at the WISE wavelengths of 12 and 23 $\mu$m, the rms
of a sine wave is 2.8 times less than the peak to peak, the amplitude goes
down to zero for pole-on orientations, and the diameter
varies like the square root of the flux.  The median and interquartile range
of the rms variation of the NEATM diameter with rotational phase
is only $7.6^{+3.0}_{-3.9}$ percent, while the median peak to peak 12 $\mu$m
flux amplitude is 0.647 mag.

It is better to evaluate the diameter at each rotational phase using the NEATM
and average these diameters than to average the fluxes and use these averages
in the NEATM.  The RMS errors for both single sample and rotationally averaged
NEATM diameters for the sample of NEOs observable by Spitzer are shown
in Figure \ref{fig:sigmarot}.  For each asteroid, 30 different pole positions and inertia
parameters $\Theta$ have been simulated, using the egg-shaped model discussed
above.

\section{Conclusion}
\label{sec:conc}

The errors in diameters computed from the NEATM for asteroid observations 
at phase angles less than $60^\circ$  are less than 10\% RMS, even for the
non-spherical shapes typical of NEOs.  For WISE, 
observing at $90^\circ$ elongation, any object with a distance 
larger than 0.6 AU will have a phase $\alpha < 60^\circ$.  
The Spitzer Space Telescope can observe at elongations between 
$85^\circ$ and $120^\circ$, so only very close passes involve $\alpha > 60^\circ$.
The error evaluated in this paper only includes the errors due to not knowing 
the thermal inertia and pole orientation of an asteroid.  There will be additional 
errors due to uncertainties in the true emissivity of the asteroid surface,
but these errors should be small, since the emissivities in the thermal 
infrared are quite close to the maximum possible value of 1.0.   
Diameters from the NEATM do not depend on the assumed albedo so there 
is no additional error from albedos.
WISE will obtain $\approx 10$ observations of each asteroid spread 
over 30 hours, and will thus get good
sampling of asteroid lightcurves, which reduces the errors associated with 
non-spherical shapes.   WISE will be sensitive enough to measure hundreds 
of thousands of asteroids, and fitting the WISE 12 \& 23 $\mu$m fluxes using 
the NEATM will provide reasonably good diameter limits for a large sample 
of asteroids.

\section{Acknowledgements}
Josh Emery and Alan Harris (DLR) provided useful comments and suggestions.
Research on WISE at UCLA is supported by the Astrophysics Division of the NASA
Science Mission Directorate.

\end{document}